\documentclass[prl,twocolumn]{revtex4}
\usepackage{graphicx}
\usepackage{amsmath,bbm}
\usepackage{amssymb,bm}

\newcommand\pd{\partial}

\newcommand\paper{Letter}
\newcommand\p{{\bm{p}}}
\newcommand\la{\langle}
\newcommand\ra{\rangle}
\newcommand\xis{\xi}
\newcommand\sect[1]{\emph{#1}\ ---}
\newcommand\ssect[1]{\emph{#1}\ ---}

\begin{document}

\author{M.~A.~Stephanov}
\affiliation{Department of Physics, University of Illinois, Chicago, 
Illinois 60607, USA}

\title{Non-Gaussian fluctuations near 
the QCD critical point}


\begin{abstract}
  We study the effect of the QCD critical point on non-Gaussian moments
  (cumulants) of fluctuations of experimental observables in heavy-ion
  collisions.  We find that these moments are very sensitive to the
  proximity of the critical point, as measured by the magnitude of the
  correlation length $\xi$. For example, the cubic central moment
  %(skewness)
  of multiplicity $\la (\delta N)^3\ra \sim \xis^{4.5}$ and
  the quartic cumulant 
  %(kurtosis)
  $\la (\delta N)^4\ra_c \sim
  \xis^7$. We estimate the magnitude of critical point contributions
  to non-Gaussian  fluctuations of pion and proton multiplicities.

\end{abstract}

\maketitle

\sect{Introduction}
Mapping the QCD phase diagram 
%as a function of temperature $T$ and
%baryochemical potential $\mu_B$ 
is one of the fundamental goals of heavy-ion collision
experiments. QCD critical point is a distinct singular feature of the
phase diagram, the existence of which is a ubiquitous property of QCD
models \cite{models} based on chiral dynamics. Locating the point
using first-principle lattice calculations is a formidable challenge
and, while recent progress and results are encouraging, much work
needs to be done to understand and constrain systematic
errors~\cite{lattice}.  If the critical point is located in the region
accessible to heavy-ion collision experiments it can be discovered
experimentally. The search for the critical point is planned at the
Relativistic Heavy Ion Collider (RHIC) at BNL, the Super Proton
Synchrotron (SPS) at CERN and the future Facility for Antiproton and
Ion Research (FAIR) at GSI.

This \paper\ focuses on the experimental observables needed to locate
the critical point in heavy-ion collisions. Locating the point
requires a scan of the phase diagram, by varying the initial collision
energy $\sqrt s$. The characteristic signature is the non-monotonous
behavior, as a function of $\sqrt s$, of the experimental observables
sensitive to the proximity of the critical point to the point where
freeze-out occurs for a given $\sqrt s$.~\cite{SRS1,Stephanov:1999zu}

The most characteristic feature of a critical point is increase and
divergence of fluctuations. Most fluctuation measures discussed
to-date can be related to {\em quadratic} variances of event-by-event
observables, such as particle multiplicities, net charge, baryon
number, particle ratios, or mean transverse momentum in the
event~\cite{Jeon:2003gk}. Typically, the singular contribution to
these variances induced by the proximity of the critical point is
proportional to approximately $\xis^2$, where $\xis$ is the
correlation length which, in the idealized thermodynamic limit, would
diverge at the critical point~\cite{Stephanov:1999zu}. The magnitude
of $\xis$ is limited trivially by the system size, but most
stringently by the finite-time effects due to critical slowing down
\cite{Stephanov:1999zu,Berdnikov}.  The observation~\cite{Berdnikov}
that the correlation length may reach at most the value of $2-3$ fm,
compared to its ``natural'' value of 1 fm, may make discovering the
critical non-monotonous contribution to such fluctuation measures a
challenging task, if the measures depend on $\xis$ too weakly.

In this \paper\ we point out that higher,
{\em non-Gaussian}, moments of the fluctuations are significantly more
sensitive to the proximity of the critical point than the commonly
employed measures based on quadratic moments. We explore this
observation quantitatively and calculate the critical point
contribution to selected fluctuation observables.

\ssect{Illustration}
The point can be illustrated using a description of the
fluctuations based on the probability distribution of an order
parameter field -- any field which, by quantum numbers, 
can mix with the critical mode $\sigma$ -- the
mode developing infinite correlation length at the critical point.

We choose the maximum of the probability distribution to be at
$\sigma=0$. The probability distribution $P[\sigma]$ can be written as
\begin{equation}
  \label{eq:P-Omega-sigma}
  P[\sigma] \sim \exp\left\{ - \Omega[\sigma]/T\right\},
\end{equation}
where $\Omega$ is the effective action (free energy) functional for
the field $\sigma$, which can be expanded in powers of $\sigma$ as
well as in the gradients:
\begin{equation}
  \label{eq:Omega-sigma}
  \Omega%[\sigma] 
= \int\!d^3x\,\left[
\frac1{2}{(\bm\nabla\sigma)^2} +
\frac{m_\sigma^2}2 \sigma^2 
+ \frac{\lambda_3}{3}\sigma^3
+ \frac{\lambda_4}{4}\sigma^4 + \ldots
\right]
\,.
\end{equation}
Near the critical point $m_\sigma\ll T$, so the mode $\sigma$ can be
treated as a classical field.

Calculating 2-point correlator $\la\sigma(\bm x)\sigma(0)\ra$ we
find that the correlation length $\xis=m_\sigma^{-1}$.
For correlation functions of the zero momentum mode $\sigma_0\equiv
\int\!d^3x\,\sigma(x)/V$ 
we find 
%(using
%standard notation for cumulants $\kappa_n(x)$):
%, to leading order in the limit $V\to\infty$ and $\xis\to\infty$:
\begin{equation}
  \label{eq:sigma-moments}
  \begin{split}
  &  \kappa_2=\la \sigma_0^2 \ra = \frac TV\,\xis^2\,;
%\\
%  &
\qquad
\kappa_3=\la \sigma_0^3 \ra = \frac{2 \lambda_3 T}V\, \xis^6\,;
\\
  &\kappa_4=\la \sigma_0^4 \ra_c
\equiv \la \sigma_0^4 \ra - \la \sigma_0^2 \ra^2
= \frac{6T}V [2(\lambda_3\xis)^{2} - \lambda_4]\, \xis^8\,.   
  \end{split}
\end{equation} 
The critical point is characterized by
$\xis\to\infty$. The central
observation in this \paper\ is that the higher moments (cumulants) $\kappa_3$ and
$\kappa_4$ diverge with $\xis$ much faster than the quadratic moment $\kappa_2$.
\footnote{Strictly speaking, the correlation functions
  scale slightly differently than Eqs.~(\ref{eq:sigma-moments}) might
  suggest, e.g., $\la \sigma_0^2 \ra \sim \xis^{2-\eta}$. Since the
  anomalous dimension $\eta\approx 0.04$ is very small, the difference
  between the actual asymptotic scaling and the scaling in
  Eq.~(\ref{eq:sigma-moments}) becomes discernible only for very large
  values of $\xis$, irrelevant in the context of this study. More
  importantly, the parameters $\lambda_3$ and $\lambda_4$ also scale
  with $\xi$ (see Eq.~(\ref{eq:lambda-tilde})).} 

Of course, the fluctuations of the critical mode are not measured
directly in heavy-ion collision experiments. These fluctuations do,
however, influence fluctuations of multiplicities, momentum
distributions, ratios, etc.\ of observed particles, such as pions or
protons, to which the critical mode
couples~\cite{Stephanov:1999zu}. The purpose of this \paper\ is to
determine the magnitude of these effects.

\sect{Critical contribution to experimental observables}
We shall now estimate the effect of the critical point
fluctuations on the observables such as the pion multiplicity fluctuations.
Using similar approach, it should be straightforward to construct
corresponding estimates for such observables as charge, proton number,
transverse momentum fluctuations, etc., as well as to take into
account acceptance cuts.

We shall focus on the most singular contribution, proportional to a
power of the correlation length $\xis$. This contribution can be found
using an intuitive picture described in Ref.~\cite{Stephanov:1999zu}.
In this picture one considers a joint probability distribution for the
occupation numbers $n_{\p}$ of observed particles (such as pions)
together with the value of the the critical mode field $\sigma$ (more
precisely, its zero-momentum mode $\sigma_0$), the latter treated as
classical. Due to coupling of the critical mode of the type
$\sigma\pi\pi$, the fluctuations of the occupation numbers receive
additional contribution, proportional to the corresponding correlation
functions (moments) of the fluctuations of $\sigma_0$ given by
Eq.~(\ref{eq:sigma-moments}). In this \paper, however, it will be more
convenient to use instead more formal diagrammatic method
developed in Ref.~\cite{Stephanov:2001zj}.

\begin{figure}
  \centering
  \includegraphics[height=7em]{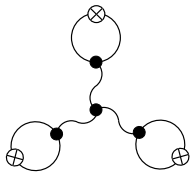}
  \caption{Diagrammatic representation of the contribution to the
    three-particle correlator from the critical mode $\sigma$. Wavy
    lines represent propagators of the $\sigma$ field, each
    contributing factor $1/m_\sigma^2$, crossed circles represent
    insertions of $\delta n_\p$ into the correlator
    Eq.~(\ref{eq:dn3-lambda}) -- see Ref.\cite{Stephanov:2001zj} for
    details.}
  \label{fig:3x}
\end{figure}

\ssect{Cubic cumulant}% (skewness)}
%\label{sec:cubic}
%
The 3-particle correlator receives the following most singular
contribution from the $\sigma$ fluctuations, given by the
diagram in Fig.~\ref{fig:3x}:
\begin{equation}
  \label{eq:dn3-lambda}
  \la \delta n_{\p_1} \delta n_{\p_2} \delta n_{\p_3}
  \ra_\sigma
=
\frac{2\lambda_3}{V^2T} \left(\frac{G}{m_\sigma^2}\right)^3
\frac{v^2_{\p_1}}{\omega_{\p_1}}
\frac{v^2_{\p_2}}{\omega_{\p_2}}
\frac{v^2_{\p_3}}{\omega_{\p_3}}
\end{equation}
where, subscript $\sigma$ indicates that only contribution of the
critical mode is considered and, as in
Refs.~\cite{Stephanov:2001zj,Stephanov:1999zu}, we denoted
$\sigma\pi\pi$ coupling by $G$ and introduced a short-hand notation
for the variance of the occupation number distribution: $v_\p^2=\bar
n_\p(1\pm \bar n_\p)$, where the ``$+$'' is for the Bose particles.

Since the total multiplicity is just the sum of all occupation numbers
and thus
\begin{equation}
  \label{eq:total-multiplicity}
  \delta N=\sum_{\p} \delta n_{\p},
\end{equation}
the cubic moment of the pion multiplicity distribution is given by
\begin{equation}
  \label{eq:dN3-lambda3}
 \la (\delta N)^3
\ra
=
V^3 \int_{\p1} \int_{\p2} \int_{\p3}    
\la \delta n_{\p_1} \delta n_{\p_2} \delta n_{\p_3} \ra
%\frac{2\lambda_3V}{T} \left(\frac{G}{m_\sigma^2}\right)^3 
%\left(\int \frac{d^3\p}{(2\pi)^3}\frac{v^2_{\p}}{\omega_{\p}}\right)^3
\end{equation}
where $\int_\p \equiv \int {d^3\p}/{(2\pi)^3}$. Since  $\la (\delta N)^3\ra$
scales as $V^1$ it is convenient to normalize it by the mean total
multiplicity $\bar N$ which scales similarly.  Thus we define
\begin{equation}
  \label{eq:omega_3-def}
  \omega_3(N)\equiv 
\frac{ \la (\delta N)^3 \ra}{\bar N}
\end{equation}
and find
\begin{equation}
  \label{eq:dN3-N-lambda3-tilde}
\omega_3(N)_\sigma
%\equiv 
%\frac{ \la (\delta N)^3 \ra_\sigma}{\bar N}
=
\frac{2\lambda_3}{T} \frac{G^3}{m_\sigma^6} 
\left(\int_\p \frac{v^2_{\p}}{\omega_{\p}}\right)^3
\left(\int_\p \bar n_\p\right)^{-1}.
\end{equation}

\begin{figure}
  \centering
  \raisebox{3em}{$3\times$}
  \includegraphics[height=7em]{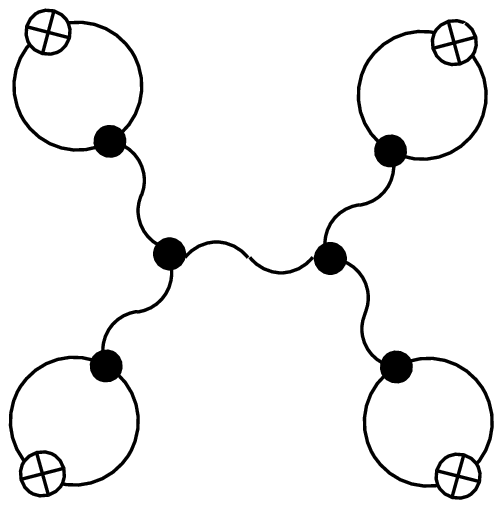}
  \raisebox{3em}{+}
  \raisebox{0.5em}{\includegraphics[height=6.4em]{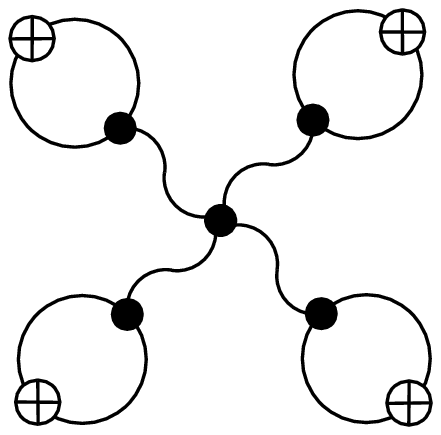}}
  \caption{Diagrammatic representation of the critical mode
    contribution to the connected four-particle correlator. Same notations as in Fig.~\ref{fig:3x}.}
  \label{fig:4x}
\end{figure}

\ssect{Quartic cumulant}% (kurtosis)}
%\label{Fourth cumulant}
%
The leading contribution to the connected 4-particle correlator
is given by the sum of two types of diagrams in Fig.~\ref{fig:4x}:
\begin{equation}
  \label{eq:dn4-lambda}
  \begin{split}
  &\la \delta n_{\p_1} \delta n_{\p_2} \delta n_{\p_3} \delta n_{\p_4}
  \ra_{c,\sigma} 
\\
&=
\frac6{V^3T}{\left[2\left(\frac{\lambda_3}{m_\sigma}\right)^2-\lambda_4\right]} \left(\frac{G}{m_\sigma^2}\right)^4
\frac{v^2_{\p_1}}{\omega_{\p_1}}
\frac{v^2_{\p_2}}{\omega_{\p_2}}
\frac{v^2_{\p_3}}{\omega_{\p_3}}
\frac{v^2_{\p_4}}{\omega_{\p_4}}
  \end{split}
\end{equation}

The quartic cumulant of multiplicity fluctuations is given by
\begin{equation}
  \label{eq:dN4-dn4}
  \begin{split}
% \kappa_4(\delta N) 
%=
\la (\delta N)^4
\ra_c
&
%= \sum_{\p_1\p_2\p_3\p_4}
=V^4\int_{\p_1\p_2\p_3\p_4}
\la \delta n_{\p_1} \delta n_{\p_2} \delta n_{\p_3} \delta n_{\p_4}
  \ra_c
%\\ &
%=
%V^4\int
%\frac{d^3\p_1}{(2\pi)^3}
%\frac{d^3\p_2}{(2\pi)^3}
%\frac{d^3\p_3}{(2\pi)^3}
%\frac{d^3\p_4}{(2\pi)^3}
%\la \delta n_{\p_1} \delta n_{\p_2} \delta n_{\p_3} \delta n_{\p_4}
%  \ra_c
%\\ &
% = 
% \frac{6(2(\lambda_3/m_\sigma)^2-\lambda_4) V}{T} \frac{G^4}{m_\sigma^8} 
% \left(\int \frac{d^3\p}{(2\pi)^3}\frac{v^2_{\p}}{\omega_{\p}}\right)^4
  \end{split}
\end{equation}
This cumulant also scales as $V^1$ in thermodynamic limit.
As in Eq.~(\ref{eq:omega_3-def}) we define a ratio whose
$V\to\infty$ limit is finite:
%It is convenient to consider a ratio which would approach a finite value as
%$V\to\infty$, such as, e.g.,
$\omega_4(N)\equiv \la (\delta N)^4\ra_c/\bar N$, and find
\begin{equation}
  \label{eq:dN4/N}
  \omega_4(N)_\sigma
%\equiv 
%\frac{\la (\delta N)^4\ra_c}{\bar N} 
=
%\frac{6(2(\lambda_3/m_\sigma)^2-\lambda_4) }{T} 
\frac6{T}{\left[2\frac{\lambda_3^2}{m_\sigma^2}-\lambda_4\right]}
\frac{G^4}{m_\sigma^8} 
\left(\int_\p\frac{v^2_{\p}}{\omega_{\p}}\right)^4
\left(\int_\p \bar n_\p\right)^{-1} .
\end{equation}

\ssect{Estimate of the effect}
In order to estimate the magnitude of the effect, we need to estimate the
values of the coupling constants $\lambda_3$ and $\lambda_4$. The main
uncertainty in the estimate will come, however, from the uncertainty
of the value of $G$, which enters in a large power. This constant is
known only roughly -- the estimate was made in
Ref.\cite{Stephanov:1999zu}. Therefore, a crude estimate
for $\lambda$'s suffices. 

\begin{figure}
  \centering
  \includegraphics[height=17em]{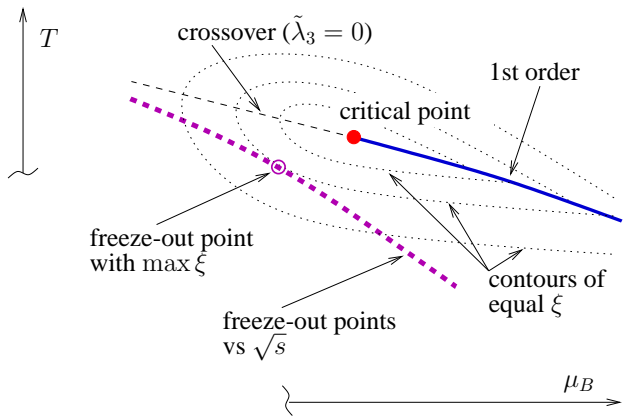}
  \caption{Illustration of the possible relative position of the critical point and the locations of the freeze-out points for different values of the initial collision energy $\sqrt s$ in a heavy-ion collision.}
  \label{fig:freezeout}
\end{figure}

Near the critical point both $\lambda_3$ and
$\lambda_4$ vanish with a power of $\xi$ given by (neglecting $\eta\ll1$):
\begin{equation}
  \label{eq:lambda-tilde}
  \lambda_3=\tilde\lambda_3T\cdot (T\xis)^{-3/2},
\qquad \mbox{and} \quad
 \lambda_4=\tilde \lambda_4 \cdot(T\xis)^{-1},
\end{equation}
where dimensionless couplings  $\tilde\lambda_3$ and $\tilde\lambda_4$ are universal, 
and for the Ising universality class they
have been measured (see, e.g., Ref.\cite{Tsypin} for a review). $\tilde\lambda_3$
varies from 0 to about 8 depending on the direction of
approach to the critical point (crossover or first-order transition
side). The coupling $\tilde\lambda_4$ varies from about $4$ to about $20$.
Since the freeze-out occurs somewhere between these two extremes (as
illustrated in Fig.~\ref{fig:freezeout}) we
shall pick some mid-range values for our estimates. The main point is
the strong dependence of the effect on~$\xis$.

Putting together estimates of $\lambda$, $G$ (from \cite{Stephanov:1999zu}) and 
$\xis$ (from \cite{Berdnikov}) we find for pions at $T\approx \mbox{120 MeV}$ 
(in full acceptance, for a single pion species, see also
Eq.~(\ref{eq:omega-k})):
\begin{equation}
  \label{eq:omega_3-pions}
   \omega_3(N_\pi)_\sigma
%\equiv \frac{ \la (\delta N_\pi)^3 \ra}{\bar N_\pi}
\approx 
1.\, 
\left(\frac{\tilde\lambda_3}{4.}\right)
\left(\frac G{\mbox{300 MeV}}\right)^3
\left(\frac{\xis}{\mbox{3 fm}}\right)^{9/2}
\end{equation}
\begin{equation}
  \label{eq:dN4/N-lambda}
  \omega_4(N_\pi)_\sigma
%=\frac{\la (\delta N_\pi)^4\ra_c}{\bar N_\pi} 
\approx  12. 
\left(\frac{2\tilde\lambda_3^2-\tilde\lambda_4}{50.} \right)
\left(\frac G{\mbox{300 MeV}}\right)^4
\left(\frac{\xis}{\mbox{3 fm}}\right)^7
\end{equation}
Because of large powers of the coupling $G$, which is known only
poorly, the uncertainty in this result is significantly larger than
that of similar estimates of the {\em quadratic} moments of
fluctuations \cite{Stephanov:1999zu}.

The most significant feature of this result is the strong dependence
on 
%the correlation length
 $\xis$ which makes the cubic and quartic cumulants
very sensitive signatures of the critical point.

\ssect{Another example: proton multiplicity fluctuations}
%\label{sec:protons}
%
The above analysis carries over to the proton multiplicity
fluctuations.  The fluctuations of the net proton number is a good
proxy to the baryon number fluctuations, whose magnitude, proportional
to the baryon number susceptibility, must diverge at the critical
point Ref.\cite{Hatta:2003wn}. This susceptibility, as well as
``kurtosis'', $\kappa_4/\kappa_2$, 
have been studied on the lattice using Taylor expansion
around $\mu_B=0$ and in QCD models~\cite{lattice,Gavai:2005sd}. Here,
for simplicity, we shall present the results for the proton only
multiplicity, which is easier to measure in experiments.

To adapt equation such as Eq.~(\ref{eq:dn3-lambda}) to protons, one
needs to substitute $G$ with the coupling $g$ of the critical mode
$\sigma$ to protons (times the mass $m_p$ of the proton): $G\to g
m_p$.  The estimate for this coupling can be taken from the
sigma-model to be roughly $g\approx m_p/f_\pi\approx 10.$ The variance
of the occupation number distribution is $v_\p^2=\bar n_\p(1-\bar
n_\p)$ and $\bar n_\p=(e^{(\omega_\p-\mu_B)/T}+1)^{-1}$, where $\mu_B$
is baryochemical potential. There is also a factor of
$2^{k-1}$ for $\omega_k$ (as in Eq.~(\ref{eq:omega-k}))
because of the proton spin degeneracy. Putting
this together one finds, e.g., for protons at SPS freeze-out
conditions $(T,\mu_B)\approx (168,266) \mbox{ MeV}$~\cite{BraunMunzinger:1995bp}
\begin{equation}
  \label{eq:omega_3-protons-sps}
   \omega_3(N_p)_\sigma
%\equiv \frac{ \la (\delta N_p)^3 \ra}{\bar N_p}
\approx 6.
\left(\frac{\tilde\lambda_3}{4.}\right)
\left(\frac g{10.}\right)^3
\left(\frac{\xis}{\mbox{1 fm}}\right)^{9/2}
\end{equation}
\begin{equation}
  \label{eq:dN4/N-lambda-protons}
  \omega_4(N_p)_\sigma
%=\frac{\la (\delta N_p)^4\ra_c}{\bar N_p} 
\approx
46.\ 
\left(\frac{2\tilde\lambda_3^2-\tilde\lambda_4}{50.} \right)
\left(\frac g{10.}\right)^4
\left(\frac{\xis}{\mbox{1 fm}}\right)^7
\end{equation}
Note that the effect is much larger on the proton multiplicity fluctuations,
compared to the pion multiplicity.

Similar to quadratic fluctuations~\cite{Hatta:2003wn}, the exponents
in Eqs.~(\ref{eq:omega_3-protons-sps}),~(\ref{eq:dN4/N-lambda-protons})
agree (up to $\eta\ll1$) with the critical exponents of the baryon number
cumulants dictated by scaling and universality:
\begin{equation}
  \label{eq:kappa-nb}
%   \kappa_k(\delta N_B)
%\equiv 
\la (\delta N_B)^k\ra_c
=
VT^{k-1}
\frac{\pd^k P(T,\mu_B)}{\pd\mu_B^k}
%\left(\frac{V}{T}\right)^{k-1}\!\!
%V^{-1}T^{1-k}\!
%\sim \xi^{(5k-6-k\eta)/2} .
\sim \xi^{k(5-\eta)/2-3} .
\end{equation}

\ssect{Mean transverse momentum}
From the expression for the correlators Eq.~(\ref{eq:dn3-lambda}) or
(\ref{eq:dn4-lambda}) one can similarly estimate the effect of the
critical point on other observables, for example, higher moments
of the fluctuation of mean transverse momentum $p_T$. For example, the
cubic moment $\kappa_3(\delta p_T)$ of the mean $p_T$ distribution around the
all-event mean  $\bar p_T$ can be
expressed as
\begin{equation}
  \label{eq:pt3}
  \begin{split}
      &\kappa_3(\delta p_T) 
\equiv \la (p_T - \bar p_T)^3 \ra 
%\\
%&
= 
\sum_{\p_1,\p_2,\p_3} 
([\p_1]_T - \bar p_T)
\\
&\qquad\times
([\p_2]_T - \bar p_T)([\p_3]_T - \bar p_T)
%\\
%&\qquad\qquad\qquad\times
 \la \delta n_{\p_1} \delta n_{\p_2} \delta n_{\p_3}
  \ra
  \end{split}
\end{equation}
and estimated using Eq.~(\ref{eq:dn3-lambda}). Normalizing this
variable similarly to the variable $F$ proposed in
Ref. \cite{Stephanov:1999zu} removes $N$ scaling and makes it less
sensitive to the effect of the flow (``blue shift'' of momenta):
\begin{equation}
  \label{eq:F3}
  F_3\equiv \frac{\kappa_3(p_T)}{\bar N v_{\rm inc}^{3/2}(p_T)}
\end{equation}
where $v_{\rm inc}(p_T)$ is the variance of the inclusive
(single-particle) $p_T$ distribution. We leave this to future work.

%There is also a trivial
%quantum statistics contribution to this moment.

\sect{Comments and discussion}
%\label{sec:discuss}
%
%
%Protons: is there a maximum possible value for $\omega_i(N)$?
%
%
It is worth noting that, even though the $\xis$ dependence of
$\omega_4$ is stronger, its measurement involves subtraction of two
contributions, $\la(\delta N)^4\ra-3\la(\delta N)^2\ra^2$, each of
which is order $N$ times larger than their difference, which might
dilute the signal-to-noise ratio in experimental measurement.

Since the freeze-out occurs, generically, somewhat off the crossover
line, as illustrated in Fig.~\ref{fig:freezeout}, one should expect the
critical point contribution to fluctuations to be skewed. In this
case, the deviations from the Gaussian shape are
dominated by the cubic moment,  or $\omega_3$.

What is the sign of $\omega_3(N)_\sigma$? One can anticipate it by
using the following, admittedly crude, argument. The skewness of the
distribution of the order parameter near the critical point is a
``shadow'' of a second peak. This peak corresponds to the phase on the
other side of the first-order transition line (quark-gluon plasma
phase at higher $T$ and $\mu_B$ on Fig.~\ref{fig:freezeout}).  This
phase has higher entropy and baryon number, thus fluctuations of these
quantities must be skewed toward higher values: $\omega_3>0$.  Since pion
and proton numbers are rough proxies to entropy and baryon number
respectively their skewness should be also positive. 
%The positive sign
%of $\omega_3(N_\pi)_\sigma$ also agrees with the relative sign between the
%couplings $\lambda_3$ and $G$ in a linear $\phi^4$ sigma model.

What is the ``natural'', background value one should expect for,
e.g., $\omega_3$? For a gas of classical free particles (Poisson
distribution) $\omega_3(N)=1$. Bose statistics increases this by a
factor $\omega_3(N)_{\rm BE}=\overline{(1+ n_{\p})(1+2 n_{\p})}$,
e.g., approximately $1.3 $ for
pions at $T=120 \mbox{ MeV}$.

More importantly, quantum statistics only correlates fluctuations of
particles of the same species, thus $\omega_3(N)_{\rm BE}$ is the same for all
charge (i.e., $N=N_{\pi^+}+N_{\pi^-}$) and single charge multiplicity
fluctuations.  In contrast, the critical point contribution correlates
also $\pi^+$ with $\pi^-$, thus making $\omega_3$ 4 times larger
for all charge vs single charge (for $\bar N_{\pi^+}=\bar N_{\pi^-}$) :
\begin{equation}
  \label{eq:omega-k}
  \omega_k(N_{\pi^+}+N_{\pi^-})_\sigma = 2^{k-1}\,\omega_k(N_{\pi^+})_\sigma.
\end{equation}
Eq.~(\ref{eq:omega-k}) can help separate critical point contribution
from contributions due to quantum statistics.

It is important to note that other sources may and do contribute to
the skewness and kurtosis: remnants of initial fluctuations, flow,
jets -- to name just a few obvious contributors. Identification and
evaluation of these contributions is a task well beyond this
paper. This serves to emphasize that the energy scan of the QCD phase
diagram is needed to separate such background contributions from the
genuine critical point effect, the latter being non-monotonous
function of the initial collision energy $\sqrt s$ as the critical
point is approached and then passed. The fact that non-Gaussian moments
have stronger dependence on $\xis$ than, e.g., quadratic moments,
makes those higher moments more sensitive signatures of the critical
point.

\acknowledgments

This work is a direct consequence of discussions at the INT program
``The QCD Critical Point''. The author is grateful to the
participants, and especially to P.~de~Forcrand, S.~Gupta, F.~Karsch,
V.~Koch, J.~Randrup, K.~Rajagopal, K.~Redlich, D.~Son and M.~Tsypin for
stimulating and useful discussions. The hospitality of the Institute
of the Nuclear Theory at the University of Washington and partial
support of the Department of Energy during the completion of this work
is gratefully acknowledged. This work is supported by the DOE grant
No.\ DE-FG0201ER41195.

\end{document}